\documentclass[a4paper]{jpconf}

\usepackage{graphicx}
\usepackage{amsmath}
\usepackage{amssymb}

\begin{document}
\title{Stabilization of direct numerical simulation for finite truncations of circular cumulant expansions}

\author{I V Tyulkina$^{1,2}$, D S Goldobin$^{1,2}$ and A Pikovsky$^{3,4}$}
\address{$^1$Institute of Continuous Media Mechanics UB RAS, 1 Akademika Koroleva street,\\
         Perm 614013, Russia}
\address{$^2$Department of Theoretical Physics, Perm State University, 15 Bukireva street,\\
         Perm 614990, Russia}
\address{$^3$Institute for Physics and Astronomy, University of Potsdam, 24/25 Karl-Liebknecht-Strasse,
         Potsdam 14476, Germany}
\address{$^4$Department of Control Theory, Nizhny Novgorod State University,
         23 Gagarin Avenue,
         Nizhny Novgorod 606950, Russia}
\ead{irinatiulkina95@gmail.com, Denis.Goldobin@gmail.com, pikovsky@uni-potsdam.de}

\begin{abstract}
We study a numerical instability of direct simulations with truncated equation chains for the ``circular cumulant'' representation and two approaches to its suppression.
The approaches are tested for a chimera-bearing hierarchical population of coupled oscillators.
The stabilization techniques can be efficiently applied without significant effect on the natural system dynamics within a finite vicinity of the Ott--Antonsen manifold for direct numerical simulations with up to $20$ cumulants; with increasing deviation from the Ott--Antonsen manifold the stabilization becomes more problematic.
\end{abstract}

\section{Introduction}
In studies of collective phenomena, especially interesting are the cases, where a network of elements exhibits nontrivial dynamics, while the inherent dynamics of elements is simple and coupling between elements or/and an external driving of them are small. This is the case of populations of self-sustained periodic oscillators with a weak mutual coupling or/and subject to a weak common forcing. For description of such system a phase reduction proved~\cite{Winfree-1967} to be applicable and the paradigmatic Kuramoto model~\cite{Kuramoto-1975} was found to be relevant for many physical, chemical and biological systems~\cite{Kuramoto-2003,Acebron-etal-2005}. The examples include both continuous distributed parameter systems and networks of discrete elements.

Later on, Kuramoto system as well as a broad class of paradigmatic phase systems of form
\begin{equation}
\dot\varphi_k=\Omega(t)+\mathrm{Im}(2h(t)\,e^{-i\varphi_k})\,,
\qquad k=1,...,N\,,
\label{eq:phiOA}
\end{equation}
where $\Omega(t)$ and $h(t)$ are arbitrary real and complex-valued functions of time, were found to be partially integrable. Watanabe--Strogatz~\cite{Watanabe-Strogatz-1993,Watanabe-Strogatz-1994,Pikovsky-Rosenblum-2008,Marvel-Mirollo-Strogatz-2009} and Ott--Antonsen theories~\cite{Ott-Antonsen-2008} were developed for such systems. While WS theory allowed studying the dynamics of finite populations~\cite{Marvel-Strogatz-2009,Zaks-Tomov-2016,Chen-Engelbrecht-Mirollo-2017} and made a solid foundation for OA theory, the latter one, constructed for the thermodynamic limit of large populations, and especially its extension to the case of nonidentical elements~\cite{Ott-Antonsen-2009,Mirollo-2012,Pietras-Daffertshofer-2016}, found broad applications and helped to elucidate many sophisticated collective phenomena~\cite{Abrams-etal-2008,Martens-etal-2009,So-Barreto-2011,Laing-2009,Nagai-Kori-2010,Laing-2012,Laing-2014,Luke-Barreto-So-2014,Pazo-Montbrio-2014,Montbrio-Pazo-Roxin-2015,Pimenova-etal-2016,Hannay-etal-2018}.

Simultaneously, the peculiarity of the systems of type~(\ref{eq:phiOA}) necessitated construction of a perturbation theory for the OA approach, since real system can be close to (\ref{eq:phiOA}) but never perfectly possess such a form. In~\cite{Tyulkina-etal-2018}, a formalism of ``circular cumulants'' formally corresponding to the Kuramoto--Daido order parameters $a_n=\langle(e^{i\varphi})^n\rangle=N^{-1}\sum_{k=1}^{N}e^{in\varphi_k}$
was suggested. Specifically, with a moment-generating function $F(\zeta)=\sum_{j=0}^{\infty}a_j\zeta^j/j!$, one can introduce circular cumulants $\varkappa_j$ via power series of the generating function $\zeta\frac{\partial}{\partial\zeta}\ln{F(\zeta)} =\sum_{j=1}^{\infty}\varkappa_j\zeta^j$. For instance, the first three cumulants are: $\varkappa_1=a_1$, $\varkappa_2=a_2-a_1^2$, $\varkappa_3=(a_3-3a_2a_1+2a_1^3)/2$\,. This formalism turned out to be fruitful for generalization of the Ott--Antonsen theory~\cite{Tyulkina-etal-2018,Goldobin-etal-2018,Tyulkina-etal-2019}.

In particular, an important case of violation of the applicability conditions of OA theory is the one of population on phase elements
\begin{equation}
\dot\varphi_k=\Omega_k+\mathrm{Im}(2h(t)\,e^{-i\varphi_k})+\sigma\xi_k(t),
\qquad k=1,...,N
\label{eq001}
\end{equation}
with natural frequencies $\Omega_k$, which are distributed around the mean value $\Omega_0$ according to the Lorentzian distribution of half-width $\gamma$, and individual white gaussian noises of strength $\sigma$: $\langle\xi_k(t)\rangle=0$, $\langle\xi_k(t) \xi_m(t')\rangle=2\delta_{km}\delta(t-t')$.
The collective dynamics of population~(\ref{eq001}) for large $N$ is governed by the following 
system of equations for circular cumulants~\cite{Tyulkina-etal-2018}:
\begin{equation}
\dot\varkappa_j=j(i\Omega_0 -\gamma)\varkappa_{j}+h\delta_{1j} -h^\ast(j^2\varkappa_{j+1}+j\sum_{m=1}^{j}\varkappa_{j-m+1}\varkappa_{m}) -{\sigma^2}(j^{2}{\varkappa}_{j}+j\sum_{m=1}^{j-1}\varkappa_{j-m}\varkappa_{m}),
\label{eq002}
\end{equation}
where $\delta_{1j}=1$ for $j=1$ and $\delta_{1j}=0$ for $j\ne1$. For $\sigma=0$, equation system~(\ref{eq002}) 
admits a particular solution $\varkappa_1=a_1$, $\varkappa_{j\ge2}=0$; this case reproduces the OA theory.

While equations~(\ref{eq002}) are useful for analytical studies and for 
construction of perturbation theories~\cite{Tyulkina-etal-2018,Goldobin-etal-2018,Tyulkina-etal-2019}, 
direct numerical simulations with them require truncations of the infinite system. 
Truncations with more than the first two cumulants kept were revealed to 
be able to cause a numerical instability. This issue turned out to be especially 
important for chimera-bearing systems~\cite{Pikovsky-Rosenblum-2008,Abrams-etal-2008,Laing-2009}, 
where neutrally stable or weakly stable degrees of freedom appear to be rather snesitive to 
the numeric instabilities. On the other hand, chimera regimes are in focus of many 
research works (e.g.,~\cite{Omelchenko-etal-2008,Omelchenko-etal-2011}) also due 
to similarities with turbulence intermittency in fluid 
dynamics~\cite{Barkley-Tuckerman-2005,Moxey-Barkley-2010}.

In this paper we suggest two approaches to suppression of 
the numerical instability of direct simulations with truncated 
chains~(\ref{eq002}) and test them for a chimera-bering 
system~\cite{Pikovsky-Rosenblum-2008,Abrams-etal-2008} described 
in section~\ref{sec:abr}. In section~\ref{sec:diss}, the stabilization by means 
of a dissipation term for the high-order cumulants is studied. In section~\ref{sec:supr}, 
we consider the approach based on the suppression of terms causing instability. 
In section~\ref{sec:concl}, the findings are summarized.

\section{Hierarchical population of coupled oscillators}
\label{sec:abr}
Abrams et al.\ \cite{Abrams-etal-2008} studied the collective dynamics of an ensemble of hierarchically coupled oscillators, composed by two subpopulations of equal size $N$; the strength of a Kuramoto--Sakaguchi coupling between oscillators~\cite{Kuramoto-1975,Sakaguchi-Kuramoto-1986,Sakaguchi-Shinomoto-Kuramoto-1988} within each subpopulation, $(1+A)/2$, is larger than the coupling strength $(1-A)/2$ between subpopulations:
\begin{equation}
\begin{aligned}
&\dot\varphi_k=\Omega +\frac{1+A}{2N}\sum_{j=1}^N\sin(\varphi_j-\varphi_k-\alpha) +\frac{1-A}{2N}\sum _{j=1}^N\sin(\psi_j-\varphi_k-\alpha) +\sigma\xi_k(t)\,,\\
&\dot\psi_k=\Omega +\frac{1+A}{2N}\sum_{j=1}^N\sin(\psi_j-\psi_k-\alpha) +\frac{1-A}{2N}\sum_{j=1}^N\sin(\varphi_j-\psi_k-\alpha) +\sigma\eta_k(t)\,,
\end{aligned}
\label{eq:abr1}
\end{equation}
where $\alpha$ is the coupling phase shift, $\sigma$ is the intrinsic noise strength, $\xi_k$ and $\eta_k$ 
are independent white Gaussian noises. For vanishing noise, the system satisfies the 
conditions of Watanabe--Strogatz and Ott--Antonsen 
theories~\cite{Watanabe-Strogatz-1993,Watanabe-Strogatz-1994,Ott-Antonsen-2008}, 
which provided a substantial progress in understanding its 
dynamics~\cite{Pikovsky-Rosenblum-2008,Abrams-etal-2008}. The chimera states, 
where one of subpopulations is fully synchronized $\psi_1=\psi_2=...=\Psi$, while 
the other one is in a partial synchrony state, were attracting the main interest of 
researchers~\cite{Abrams-etal-2008}.

However, the intrinsic noise violates these properties, and the circular 
cumulant approach proved to be useful for the description of the noisy 
system~\cite{Tyulkina-etal-2018,Goldobin-etal-2018}. In the thermodynamic limit of large $N$, 
one can introduce the circular cumulant description, with cumulants $\varkappa_j$ for 
subpopulation $\{\varphi_k\}$ and $\kappa_j$ for $\{\psi_k\}$ (cf. equation~(\ref{eq002})):
\begin{align}
&\dot\varkappa_j=j(i\Omega_0 -\gamma)\varkappa_{j}+h_\varphi\delta_{1j} -h_\varphi^\ast(j^2\varkappa_{j+1}+j\sum_{m=1}^{j}\varkappa_{j-m+1}\varkappa_{m}) -{\sigma^2}(j^{2}{\varkappa}_{j}+j\sum_{m=1}^{j-1}\varkappa_{j-m}\varkappa_{m})\,,
\label{eq:abr-c1}
\\
&\dot\kappa_j=j(i\Omega_0 -\gamma)\kappa_{j}+h_\psi\delta_{1j} -h_\psi^\ast(j^2\kappa_{j+1}+j\sum_{m=1}^{j}\kappa_{j-m+1}\kappa_{m}) -{\sigma^2}(j^{2}{\kappa}_{j}+j\sum_{m=1}^{j-1}\kappa_{j-m}\kappa_{m})\,,
\label{eq:abr-c2}
\end{align}
where $h_\varphi=0.25[(1+A)\varkappa_1+(1-A)\kappa_1]e^{-i\alpha}$ and $h_\psi=0.25[(1+A)\kappa_1+(1-A)\varkappa_1]e^{-i\alpha}$. In the classical ``Abrams' system''~\cite{Abrams-etal-2008} the oscillator natural frequencies are identical: $\gamma=0$.
In this paper we use the Abrams' system~(\ref{eq:abr1}) as a model for testing the approaches to stabilizing numerical simulation with truncated equation chains~(\ref{eq:abr-c1}), (\ref{eq:abr-c2}).

\section{Introducing auxiliary dissipation}
\label{sec:diss}
From numerical viewpoint, eqs.~(\ref{eq002}) is a system of ODEs, which,
if properly truncated, can be solved with the Runge-Kutta method. However, for 
some systems, truncation of the system can lead to a numerical instability. In this section 
we introduce additional dissipation-like terms into the original equations,
\begin{align}
\textstyle
\dot\varkappa_{j}=j(i\Omega_0 -\gamma)\varkappa_{j}+h\delta_{j1} -h^\ast\left(j^2\varkappa_{j+1}+j\sum_{m=1}^{j}\varkappa_{j-m+1}\varkappa_{m}\right) \qquad\qquad\qquad
\nonumber\\
\textstyle
{}-{\sigma^2}\left(j^{2}{\varkappa}_{j}+j\sum_{m=1}^{j-1}\varkappa_{j-m}\varkappa_{m}\right) -G(\gamma+\sigma^2)(j-1)(j-2)2^j\varkappa_{j}\,,
\label{eq101}
\end{align}
and study their effect on the numerical stability of calculations.
The choice of such a form of the stabilizing term is guided by the following reasons:
\\
$\bullet$~The $G$-term introduces additional dissipation into the system dynamics.
Similar to natural dissipation terms, it depends on parameters  $\gamma$ and $\sigma$,
but is rather strong for large $j$.
\\
$\bullet$~Since the stabilizing term must make a stronger impact on the 
higher-order cumulants, i.e.\ grow with $j$ much faster than the $\sigma^2j^2\varkappa_j$-term, 
we adopt the $j$-dependence of the form $\propto j^2 s^j\varkappa_j$. 
As simulations with only the first or two first cumulants are always stable, 
it is reasonable to amend the $j^2$-multiplier by changing it to $(j-1)(j-2)$. 
The numerical simulations revealed that the choice of $s=2$ is optimal.

We perform numerical simulations with the following initial conditions: the 
second subpopulation is synchronized $\kappa_j=\delta_{1j}$ and the elements 
of the first subpopulation form a two-bunch distribution, which is a 
superposition of two wrapped Cauchy distributions with complex order 
parameters $b_1$ and $b_2$. In terms of moments $a_j=\langle{e^{ij\varphi}}\rangle$, the latter means
\[
a_j=q_1(b_1)^j+q_2(b_2)^j,
\]
where $q_1$ and $q_2=1-q_1$ are the fractions of elements in the first and second bunch, respectively. Terms $(b_1)^j$ and $(b_2)^j$ represent the probability density distributions of $\varphi$ corresponding to OA solutions with complex order parameters $b_1$ and $b_2$, respectively. The distance between $b_1$ and $b_2$ is a convenient parameter for controlling the proximity to the OA manifold; the case $b_1=b_2$ yields an OA solution.

\begin{figure}
\begin{center}
\begin{minipage}[b]{2.85in}
\centerline{\includegraphics[width=0.92\textwidth]%
  {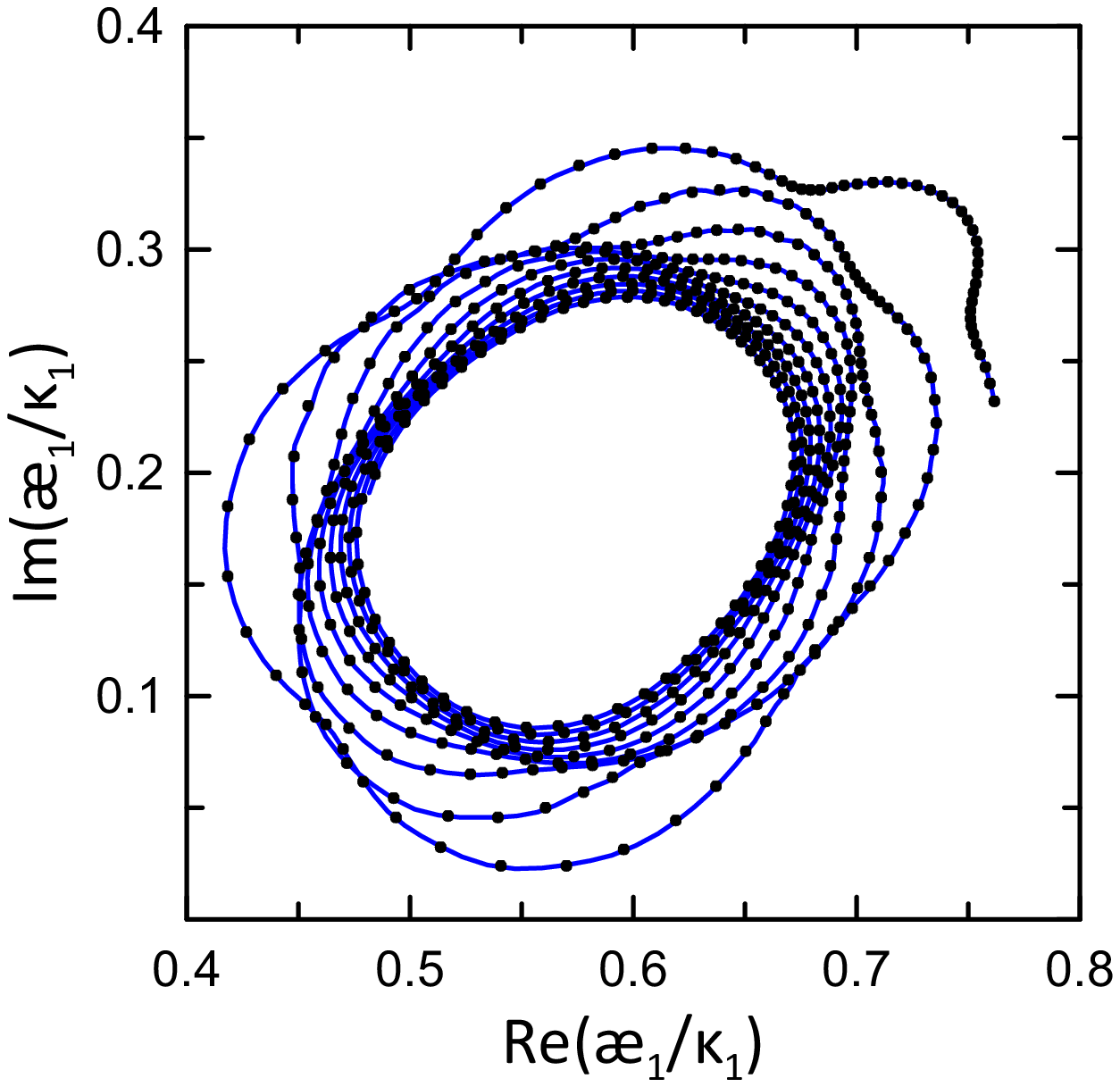}}
\caption{Trajectory of system (\ref{eq:abr-c1})--(\ref{eq:abr-c2}) for $A=0.3$, $\alpha=\pi/2-0.15$, $\gamma=0.001$, $b_{1,2}=(0.8\pm0.05)e^{i(0.3\pm0.1)}$. Black dots are the result of ``exact'' integration, blue curve is the result of numerical integration of equations~(\ref{eq101}); first 15 cumulants are used.}
\label{fig1}
\end{minipage}
\qquad
\begin{minipage}[b]{2.85in}
\centerline{\includegraphics[width=0.95\textwidth]%
  {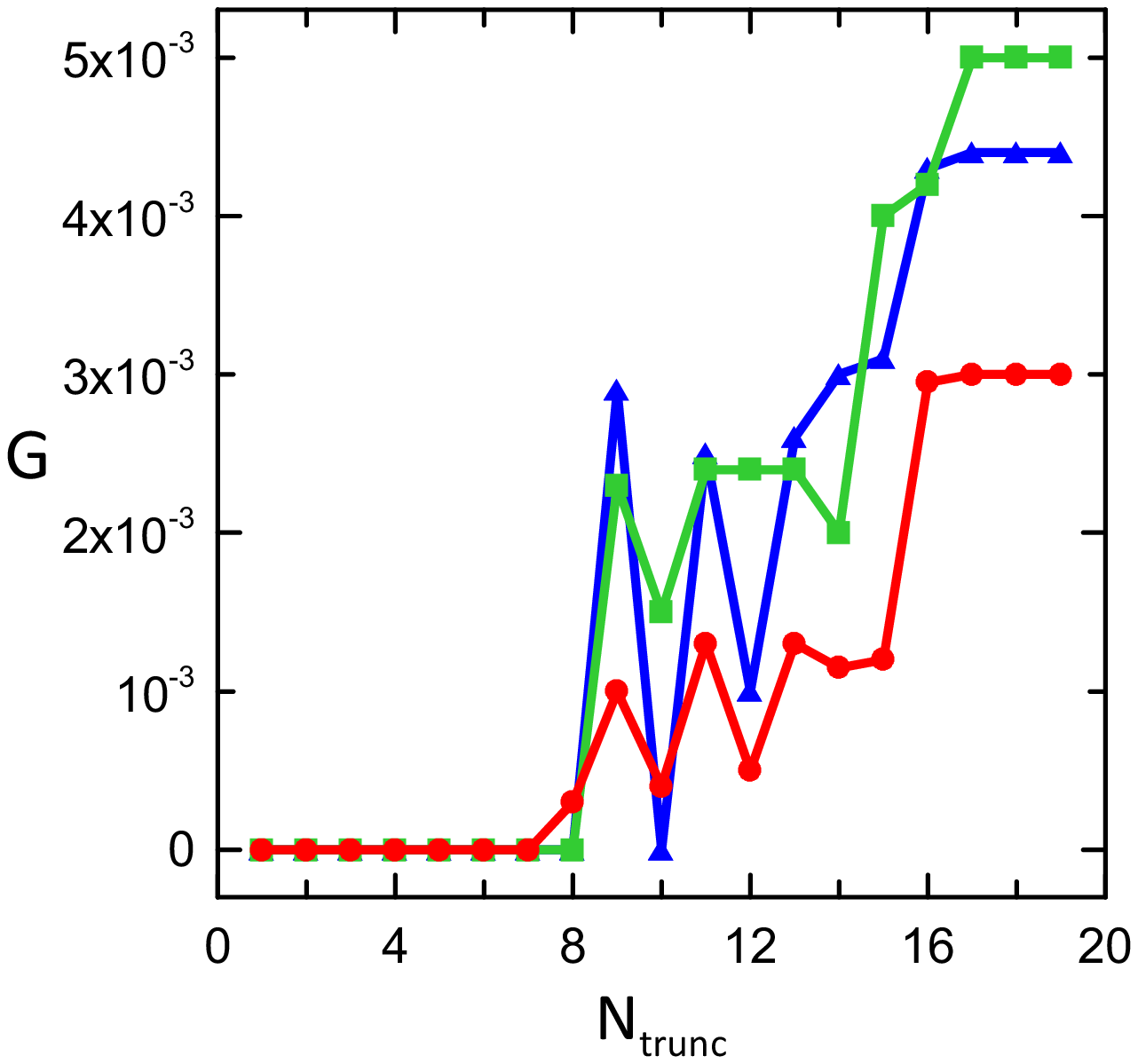}}
\caption{The critical value of the coefficient $G$ versus the number of cumulants $N_\mathrm{trunc}$ is plotted for $\gamma=0.001$. Blue triangles: $\sigma=0$, $q_1=q_2=0.5$; red circles: $\sigma=0.01$, $q_1=q_2=0.5$; green squares: $\sigma=0$, $q_1=0.4$, $q_2=0.6$.}
\label{fig2}
\end{minipage}
\end{center}
\end{figure}

Let us first consider the case of equipartition between two bunches: $q_1=q_2=0.5$. In this case, simulations can be performed accurately, due to the fact that all odd cumulants, except the first one, turn to zero~\cite{Tyulkina-etal-2019}. Indeed, one can perform simulations for $\varkappa_1$ and $\varkappa_{2n}$ with terms $\varkappa_{2n+1}=0$ wiped-out from equations for $n\ge1$; the numerical calculations are stable for any order of truncation in this case. The trajectory of the system is shown in figure~\ref{fig1}.

Figure~\ref{fig2} provides the results which can guide the choice of the values of coefficient $G$ for numerical simulations. In the graph, one can see the dependence of the critical value $G$, above which the numerical calculations are stable, on the number of cumulants kept in a truncated chain, $N_\mathrm{trunc}$. The choice of value of $G$, moderately exceeding the critical value, allows achieving compromise: the simulations are free of numerical instability, while the system trajectory is practically unaffected by the stabilizing term.

\begin{figure}
\begin{center}
\noindent
{\sf (a)}\hspace{-19pt}
\includegraphics[width=0.46\textwidth]%
  {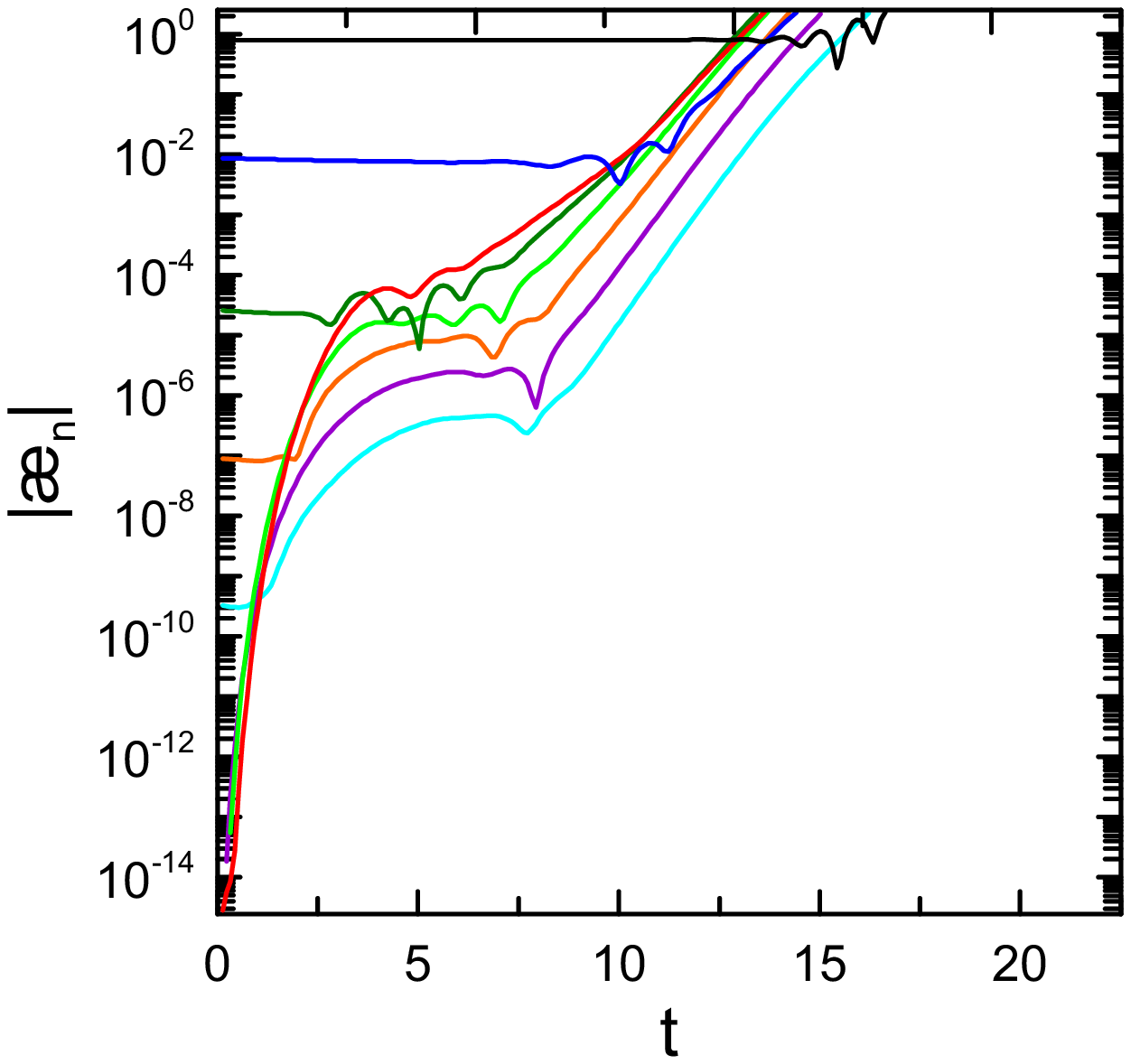}
\qquad
{\sf (b)}\hspace{-19pt}
\includegraphics[width=0.46\textwidth]%
  {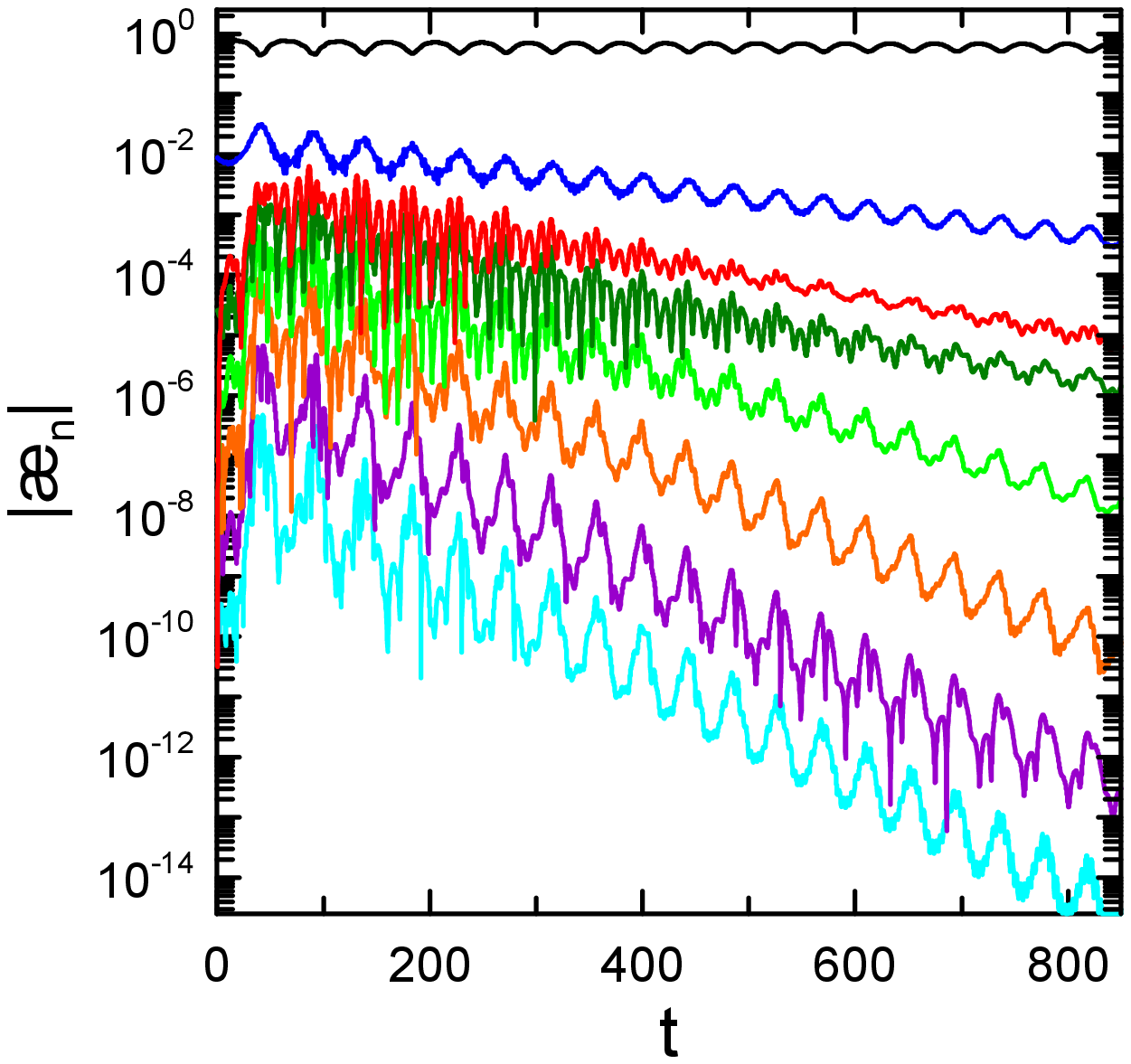}
\end{center}
\caption{Dependencies of cumulants $\varkappa_{n}$ on time $t$ (from top to bottom: from $\varkappa_1$ to $\varkappa_8$). (a)~Parameter $G$ is chosen below the stability threshold of the method (see figure~\ref{fig2}). (b)~Parameter $G$ is above the stability threshold of the method.}
\label{fig3}
\end{figure}

In figure~\ref{fig3}(a) we demonstrate the explosion of  numerical simulation due to numerical instability below the stabilization threshold; one can see that the blow-up is driven by the highest-order cumulant, the equation for evolution of which is affected by truncation of the cumulant chain. Above the threshold, in figure~\ref{fig3}(b), the evolution is regular, fast fluctuations of cumulants decay with time, and the system forms a hierarchy of smallness for cumulants as predicted theoretically~\cite{Tyulkina-etal-2018,Goldobin-etal-2018}.


Stabilization can be achieved with a minor distortion of the original system dynamics for not more than 20 cumulants and a non-large distance between the two bunches in the unsynchronized subpopulation. When using more than 20 cumulants, the value of $G$ required to stabilize the numerical calculations grows very quickly with the number of cumulants: the stabilizing term becomes significantly distorting the `true' dynamics of the original system.

\begin{figure}[!b]
\begin{center}
{\sf (a)}\hspace{-19pt}
\includegraphics[width=0.40\textwidth]%
  {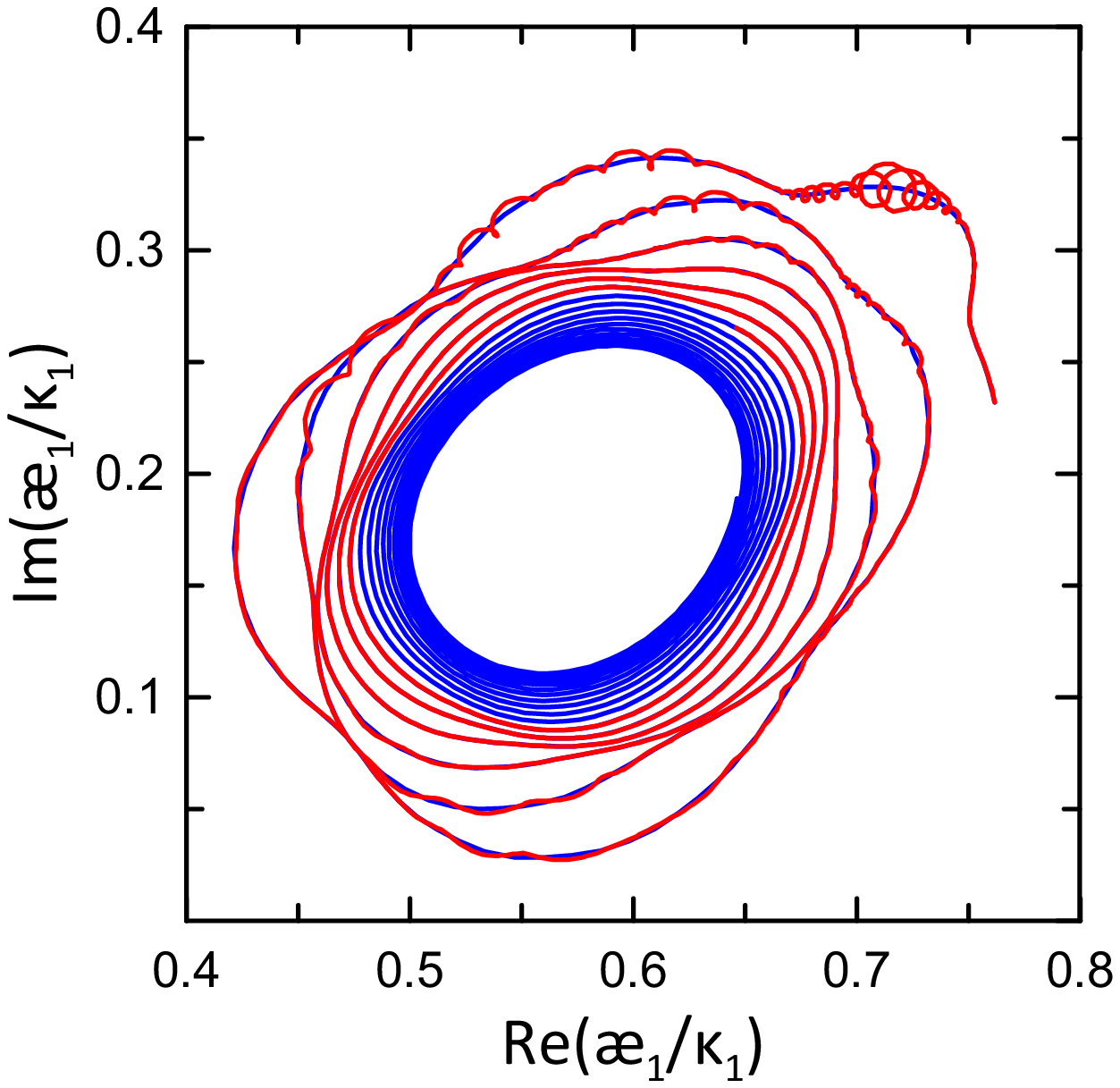}
\qquad\qquad
{\sf (b)}\hspace{-19pt}
\includegraphics[width=0.40\textwidth]%
  {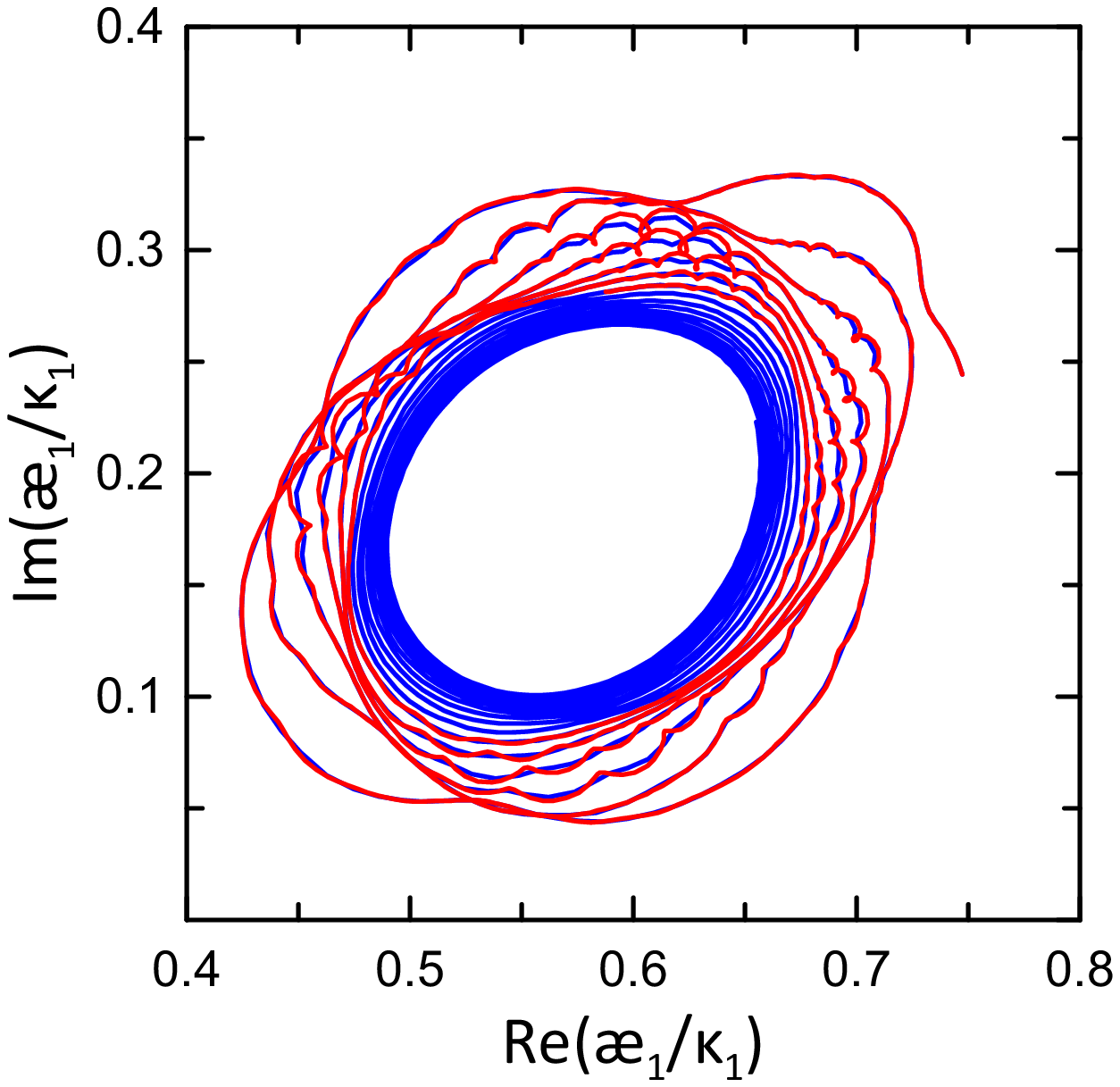}

\caption{Trajectory of the system is plotted for (a)~$q_1=q_2=0.5$, $\sigma=0.01$ and (b)~$q_1=0.4$, $q_2=0.6$, $\sigma=0$. The blue `smooth' curves: $G=0.01$; the red `curly' curves: $G$ at the stability boundary. Calculations with first 15 cumulants.}
\label{fig4}
\end{center}
\end{figure}

Let us consider the case of noise strength $\sigma=0.01$. In the unsynchronized subpopulation, the oscillators are equally distributed between two bunches, i.e.\ $q_1=q_2=0.5$. In figure~\ref{fig2} one can see that the critical value of $G$ for the noisy case (red circles) is lower than for the noise-free case (blue triangles). In figure~\ref{fig4}(a) the example of system trajectory is plotted. One can notice that the trajectory for a marginal value of $G$, although being significantly disturbed by fluctuations, does not diverge with the trajectory calculated for a larger value ​​of $G$, for which the computational fluctuations are suppressed.

Testing with the equipartition case $q_1=q_2$ at $\sigma=0$ is important, since in this case the integration of the modified equation system can be accurately performed without a numerical instability; one can check the results of numerical integration with stabilization terms against the numerical solutions for the unaltered system. However, the specificity of the equipartition solutions can also effect their numerical stability properties. Hence, an additional treatment for the case of unequal partition between two bunches is required. The numerical results for $\sigma=0$, $q_1=0.4$ and $q_2=0.6$ are presented in figures~\ref{fig2} (green squares) and \ref{fig4}(b). In figure~\ref{fig2} one can see that the characteristic critical values ​​of $G$ for equal and unequal partitions between two bunches are close. For a noisy system,  smaller values ​​of the coefficient $G$ are required, since noise makes a stabilizing effect.

From the provided results on the stability boundary of the numerical simulations, one can conclude that, for the considered initial conditions, the stabilizing term is not required if less than 8 cumulants are used. For a simulation with larger number of cumulants, a stabilizing term is required.

\begin{figure}
\begin{center}
\includegraphics[width=0.38\textwidth]%
  {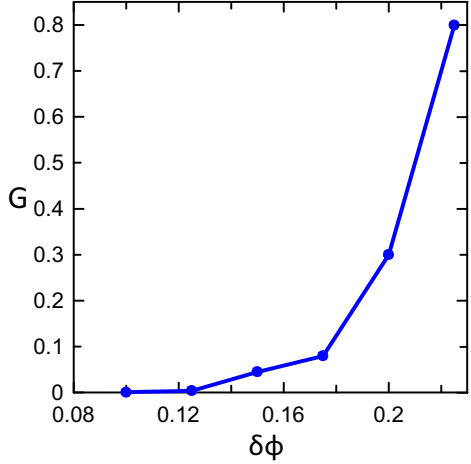}
\qquad
\begin{minipage}[b]{2.85in}
\caption{The critical value of the coeffici\-ent $G$ versus the phase distance $\delta\phi$ for $A=0.3$, $\alpha=\pi/2-0.15$, $\delta R=0.05$, $R_0=0.8$, $\Phi_0=0.3$, $\gamma=0.001$. Here $N_\mathrm{trunc}=12$ cumulants are used for calculations; $\sigma=0.01$ and the distribution of oscillators between two bunches is unequal: $q_1=0.4$ and $q_2=0.6$.}
\label{fig5}
\end{minipage}
\end{center}
\end{figure}

The efficiency of stabilization procedure depends on initial conditions. For initial conditions, the distance between the complex order parameters
\[
b_{1,2}=(R_0\pm\delta{R})e^{i(\Phi_0\pm\delta\phi)}
\]
in amplitude and phase ($\delta R$, $\delta\phi$) of two bunches is required to be non-large, otherwise the suppression of the numerical instability becomes impossible (simulation with two first cumulants is always stable). For these situations, the required value of the coefficient $G$ becomes too large and affects the natural dynamics of the system. In figure~\ref{fig5}, the dependence of the critical value of $G$ on the phase difference $\delta\phi$ is plotted for a fixed distance in amplitude between two bunches.

\section{Suppression of the terms responsible for instability}
\label{sec:supr}
An alternative approach to stabilization is the suppression of the terms which are primarily 
responsible for numerical instability in the original system. We modify one of the terms 
of the original equation by an additional exponential factor $e^{-\lambda(j-1)}$:
\begin{align}
\textstyle
\dot\varkappa_{j}=j(i\Omega_0-\gamma)\varkappa_{j}+h\delta_{j1}-h^\ast\Big(j^2\varkappa_{j+1}e^{-\lambda(j-1)}
\qquad\qquad\qquad
\nonumber\\
\textstyle
{}+j\sum_{m=1}^{j}\varkappa_{j-m+1}\varkappa_{m}\Big)
-{\sigma^2}\Big(j^{2}{\varkappa}_{j}+j\sum_{m=1}^{j-1}\varkappa_{j-m}\varkappa_{m}\Big).
\label{eq102}
\end{align}
This modification is guided by the following reason:
\\
$\bullet$~ The numerical instability is related to the $\varkappa_{j+1}$-term in $\dot\varkappa_{j}$. We modify it in such a way that for large $j$ it vanishes, while for small $j$ it is nearly unchanged, that is the modification does not affect the accuracy of calculations significantly.

We apply this stabilization procedure to the same examples as in the previous section. The second subpopulation is nearly perfectly synchronized, in the first subpopulation oscillators are partitioned between two bunches. The cases to be considered:\\
1)~Oscillators are equally distributed between two bunches, i.e.\ $q_1=q_2=0.5$; no noise, $\sigma=0$.
\\
2) Oscillators are equally distributed between two bunches, i.e.\ $q_1=q_2=0.5$, but noise is added to the system $\sigma=0.01$.\\
3) Oscillators are unequally distributed between two bunches, $q_1=0.4$ and $q_2=0.6$; $\sigma=0$.

\begin{figure}
\begin{center}
\begin{minipage}[b]{2.85in}
\centerline{\includegraphics[width=0.90\textwidth]%
  {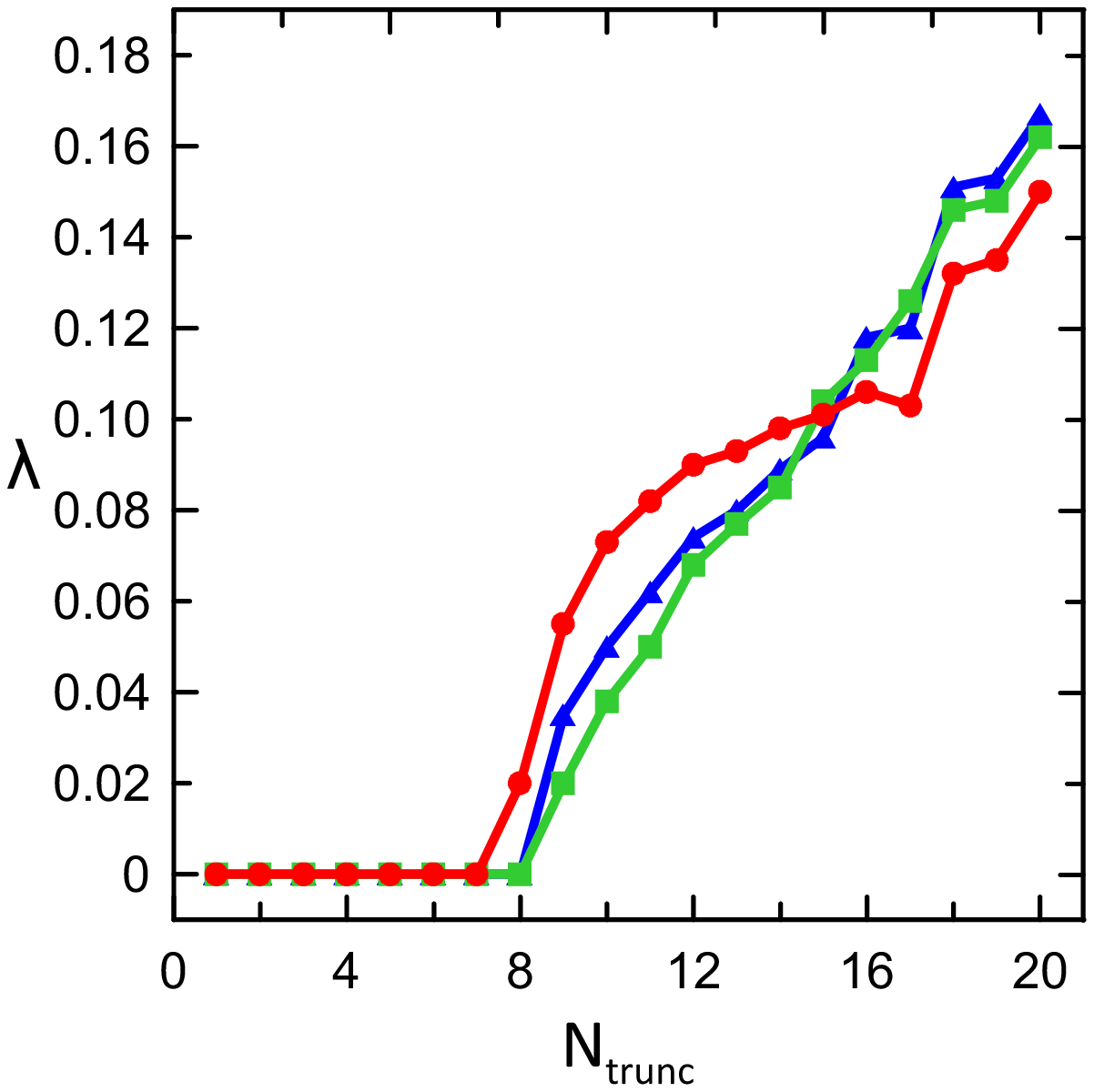}}
\caption{The critical value of the expo\-nen\-tial coefficient $\lambda$ versus the number of cumulants $N_\mathrm{trunc}$ for $\gamma=0.001$. Blue triangles: $\sigma=0$, $q_1=q_2=0.5$; red circles: $\sigma=0.01$, $q_1=q_2=0.5$; green squares: $\sigma=0$, $q_1=0.4$, $q_2=0.6$.}
\label{fig6}
\end{minipage}
\qquad
\begin{minipage}[b]{2.85in}
\centerline{\includegraphics[width=0.888\textwidth]%
  {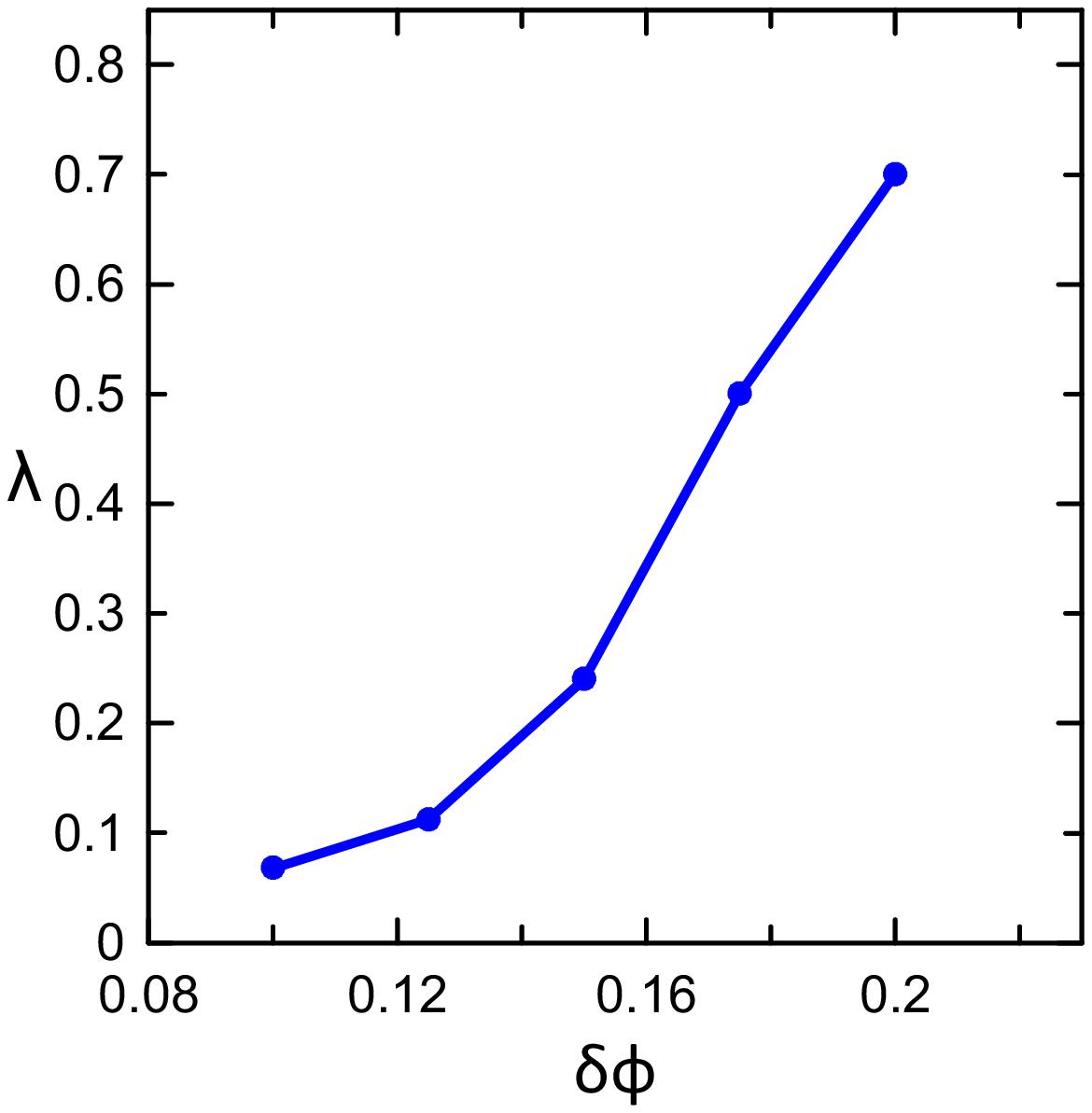}}
\caption{The critical value of the expo\-nen\-tial coefficient $\lambda$ versus the phase dif\-ference $\delta\phi$ for $A=0.3$, $\alpha=\pi/2-0.15$, $\delta R=0.05$, $R_0=0.8$, $\Phi_0=0.3$, $\gamma=0.001$, $\sigma=0.01$, unequal partition between two bunches, $q_1=0.4$, $q_2=0.6$; $N_\mathrm{trunc}=12$.}
\label{fig7}
\end{minipage}
\end{center}
\end{figure}

The minimal exponential coefficient $\lambda$ required for stabilization as a function of the number of cumulants $N_\mathrm{trunc}$ is presented in figure~\ref{fig6}. Initial conditions are the same as in the previous section (see the caption for figure~\ref{fig1}).

We also calculate the dependence of the coefficient $\lambda$ on the phase difference $\delta\phi$ between two bunches. In figure~\ref{fig7}, one can see such a dependence for the case of unequal partition of oscillators between two bunches, $q_1=0.4$ and $q_2=0.6$.


Similarly to the previous stabilization method, this method can be implemented without affecting the natural system dynamics for direct numerical simulation with no more than $20$ cumulants. The strength of the required stabilizing correction depends on the proximity to the Ott--Antonsen manifold; too far away from the OA manifold (for a large distance between two bunches), numerical simulations with large number of cumulants become impossible.

\section{Conclusion}
\label{sec:concl}
We have considered two approaches to suppress numerical instability of direct simulations 
of truncated systems of circular cumulant equations~(\ref{eq002}). With the first approach, 
an additional dissipation term for high-order cumulants is introduced (see equation~\ref{eq101}). 
With the second approach, the the coupling term term $\propto\varkappa_{j+1}$, responsible for 
instability of truncated chains, is exponentially suppressed for high-order 
cumulants (see equation~\ref{eq102}). The stability domains have been constructed 
for these stabilization methods; the dependence of stability on the initial conditions has been studied as well.

One can conclude, that the stabilization can be practically achieved 
without significant effect on the natural system dynamics for no more than 20 cumulants. 
Stabilization efficiently works in a small but finite vicinity of 
the Ott--Antonsen manifold. For a larger deviation from the OA manifold, 
direct numerical simulations with fewer number of cumulants can be stabilized.

\ack{The work has been financially supported by the Russian Science Foundation (Grant no.\ 19-42-04120).}

\end{document}